\documentclass[10pt,a4paper,twocolumn,fleqn,english]{revtex4-1}

\usepackage{bm}
\usepackage{natbib}
\usepackage{color}
\usepackage{graphicx}
\usepackage{amsmath}
\usepackage{amssymb}
\usepackage{psfrag}
\usepackage{float}
\usepackage[caption=false]{subfig}

\begin{document}

\title{A Maxwell Construction for Phase Separation in Vibrated Granular Matter?}

\author{James P.D. Clewett$^1$, Jack Wade$^2$, R. M. Bowley$^2$, Stephan Herminghaus$^1$, Michael R. Swift$^2$ and Marco G. Mazza$^1$.}

\affiliation{$^1$ Max Planck Institute for Dynamics and Selforganization, Am Fa{\ss}berg 17, 37073 G\"ottingen, Germany.}
\affiliation{$^2$ School of Physics and Astronomy, University of Nottingham, Nottingham, NG7 2RD, United Kingdom.}

\begin{abstract}
Experiments and computer simulations are carried out to investigate ordering principles in a granular gas which phase separates under vibration. The densities of the dilute and the dense phase are found to follow a lever rule. 
A Maxwell construction is found to predict both the coexisting pressure and binodal densities remarkably well, despite the fact that the pressure-volume characteristic $P(v)$ is not an isotherm.  Although the system is far from equilibrium and energy conservation is strongly violated, we derive the construction from the minimization of mechanical work and fluctuating particle currents.
\end{abstract}

\maketitle

Many-particle systems driven far from equilibrium, which occur abundantly in nature, technology, as well as in laboratory settings, often exhibit remarkable collective behaviour \cite{Meron201270, Sato03032009, RevModPhys.85.1143, Yuan2006294, Foster201188, RevModPhys.71.S396, RevModPhys.78.641, degroot}. 
In spite of the importance of such phenomena, the search for underlying principles governing their dynamics and emerging patterns is still continuing \cite{arXiv:1009.4874, Egolf07012000, 0034-4885-75-12-126001, PhysRevLett.108.210604}.  
Inspired by analogous problems in equilibrium thermodynamics, it has proven useful to study non-equilibrium steady states (NESS) which are characterized by time-independent macroscopic quantities and their fluctuations.  

A paradigmatic model system exhibiting a NESS is a driven granular gas \cite{RevModPhys.68.1259, JStatPhys.125.3.553-568, AIPCONF1.3179956, Brilliantov2010}. In its simplest form, a granular gas is a cloud of dissipative spherical particles maintained in a steady state by a continuous external drive \cite{PhysRevE.70.050301,NaturePhys.4.2008}. Recent work has demonstrated that confined grains driven by a periodic external force can separate into a dilute phase and a dense phase via spinodal decomposition \cite{PhysRevE.78.061301, PhysRevLett.109.228002}, and exhibit behavior similar to the phase separation in a van der Waals gas \cite{PhysRevLett.89.044301,Soto2003,PhysRevE.75.061304, PhysRevLett.107.048002, PhysRevLett.109.228002}. In an equilibrium fluid the pressure at coexistence, $P^*$, as well as the binodal densities $\phi_g$ (gas) and   $\phi_l$ (liquid) are determined by the requirements of mechanical and  thermal equilibrium and by the minimization of the appropriate free energy. These constraints give rise to the Maxwell construction on the non-monotonic pressure-volume isotherm $P(v)$ \cite{Maxwell1875}.  This equal-areas construction uniquely identifies $P^*=P_{EA}$, the equal-areas pressure.  

In a granular gas, the degree of dissipation is quantified by the restitution coeffitient, $\varepsilon \leq 1$, which denotes the ration of the relative momenta after/before impact of two particles. Clustering is observed whenever $\varepsilon < 1$. If $\varepsilon$ is close to unity (elastic limit), it is not surprising that the overall clustering behaviour can be described by concepts borrowed from equilibrium statistical physics \cite{PhysRevLett.89.044301,Soto2003,PhysRevE.70.051310,PhysRevE.75.061304}. However, as the system is taken far away from equilibrium, its behaviour is expected to differ qualitatively from its equilibrium counterparts. In particular, a Maxwell construction, which directly derives from the minimization of a free energy functional, is not expected to hold. Surprisingly, we find that an equal-areas rule (Maxwell construction) quite accurately predicts the coexisting binodal densities and the pressure for the liquid-gas phase separation to a few percent, even if the system is well remote from the elastic limit (with $\varepsilon$ down to 0.65). We demonstrate that this results from the minimization of residual mechanical work associated with the fluctuations in the system.

Our experimental apparatus is very similar to that used previously \cite{PhysRevLett.107.048002}.  Glass particles were confined  between parallel horizontal plates in a long, thin cell. 
The particles were sieved and selected under a microscope to obtain a sample of approximately monodisperse spheres with diameter $d$ = 610 $\mu$m. The cell was constructed from a lower plate of 3 mm thick, anodized aluminum and a top plate of 3 mm thick glass. The plates were separated by $10$ mm high aluminum walls which also confined the particles horizontally such that the internal length, width and height of the cell were $280$ mm, $10$ mm, and $10$ mm, respectively.  The cell was driven sinusoidally in the vertical direction, with variable amplitude, $A$. The driving frequency was kept fixed at $60$ Hz. The mean density is defined as $\bar{\phi}=Nv_p/V$, where $N$ is the number of particles, $v_p$ is the volume of a single particle and $V$ is the internal volume of the cell. Care was taken to ensure that the cell was level prior to each experimental run.

\begin{figure}[t]
 \begin{center}
  \includegraphics[clip,width=1.0\columnwidth]{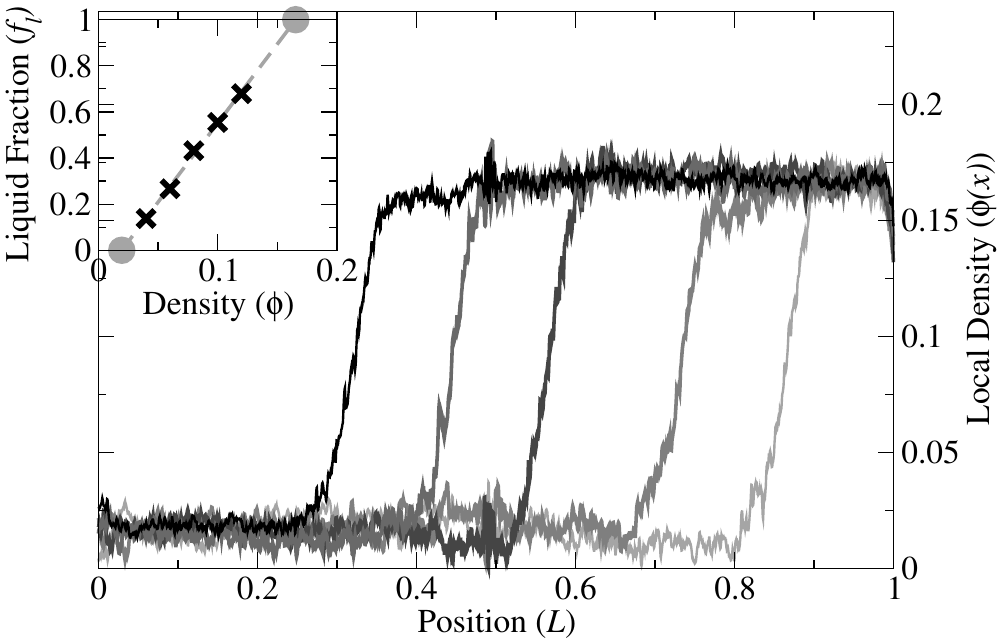}
  \caption{ Main panel: the mean grey level from photographs of the experiment. Data are shown for densities in the range $0.04\leq\bar{\phi}\leq0.12$, increasing from right to left.  The driving amplitude is $A$=2.1$d$.
  The local density, $\phi(x)$, is inferred from the average gray level in top-view photographs of the system, and scaled such that the peak values match the values predicted by the lever rule plot by extrapolation to $f_l$=0  and $f_l$=1.
  Inset:  The crosses mark the interface position obtained from the experiment. The dashed line is linear fits to the data.  Zero liquid fraction corresponds to gas density, $\phi_g$, unity liquid fraction represents liquid density, $\phi_l$ (circles).}
\label{fig:lever-rule}
 \end{center}
\end{figure}

Density profiles measured in experiments using a long cell are shown in the main panel of Fig.~\ref{fig:lever-rule}, the corresponding interface positions are shown in the inset. As $\bar{\phi}$ increases, the volume of the liquid phase increases, moving the interface to the left. The densities $\phi_l$ and $\phi_g$ appear to be independent of $\bar{\phi}$.  In the inset the linear fit demonstrates that the system  obeys a lever rule, $\bar{\phi}=\phi_lf_l+\phi_gf_g$, where  $f_l$ is the liquid fraction and $f_g$ is the gas fraction.  The lever-rule confirms that there is an intrinsic mechanism which selects the liquid and gas densities as intensive quantities of the system. In thermodynamic equilibrium, the steady state is defined by maximizing the entropy of the system subject to the conservation of energy and momentum. In the NESS, energy conservation is violated due to the external driving and the dissipation of energy into internal degrees of freedom of the particles. This invites the question: what sets the coexisting densities in this NESS?

In addition to our experiments, we have also carried out time-driven molecular dynamics simulations. The simulations have previously been shown to accurately capture the physics of the system under study \cite{PhysRevLett.107.048002, PhysRevLett.109.228002}.  The particles are modelled as monodisperse soft-spheres with diameter $d$ = 610 $\mu$m. Dissipation is included by a normal coefficient of restitution, $\varepsilon$ (implemented using a linear-spring and dash-pot damping) \cite{schwager-GM-2007}. The effects of tangential forces and rotational degrees of freedom are neglected because these have been shown to have minor impact on the physics of the system \cite{PhysRevE.52.4442, GranMatt.1434-5021, PhysRevLett.108.018001}. The simulated cell has dimensions $460d\times20d\times16.4d$, thus closely resembling the experimental system. Reflecting boundary conditions on the short walls of the cell are used in order to study a single interface between the two coexisting phases.  On the long walls, periodic  boundary conditions are employed. The results presented in this paper are for $\varepsilon$ = 0.8, matching well with the experiment, but similar results are found for the rather wide range of $0.65\leqslant\varepsilon\leqslant0.95$, that is, all coefficients of restitution for which the liquid-gas phase separation can be uniquely identified.
 
Our simulations allow us to determine quantities that are not readily available experimentally, for example the pressure within the granular gas.  We define the pressure in the homogeneous regions to be $P\equiv\tfrac{1}{2}(P_{xx}+P_{yy})$, where $P_{xx}$ and $P_{yy}$ are the horizontal components of the pressure tensor~\cite{allen1987}.  To ensure we are in the steady state we first relax the system for ten seconds of simulated time. Thereafter pressure is averaged both spatially and over ten distinct initial configurations, for ten seconds each. To obtain the local pressure, the spatial average is additionally binned into boxes of size 5$d$.

By simulating small sample cells with horizontal dimensions less than the liquid-gas interface width, phase separation can be suppressed. In this way the pressure can be calculated as a function of homogeneous quantities even under conditions for which a large system would phase separate.  In our simulations, periodic boundary conditions are used in the horizontal directions.  This method is a common practice for granular gases \cite{PhysRevLett.89.044301, PhysRevLett.107.048002, PhysRevE.65.021302, PhysRevE.66.021306,PhysRevE.70.031302}, however, recent work questions whether the non-monotonic pressure-volume curves obtained represent the equation of state for the system, or merely reflect finite size effects \cite{Takahashi1998128,AJP.1.4754020}.  Crucially we find that, for sufficiently {\em small} cells, the calculated pressure is not a function of the system size.

\begin{figure}[t]
\begin{center}
\includegraphics[clip,width=1.0\columnwidth]{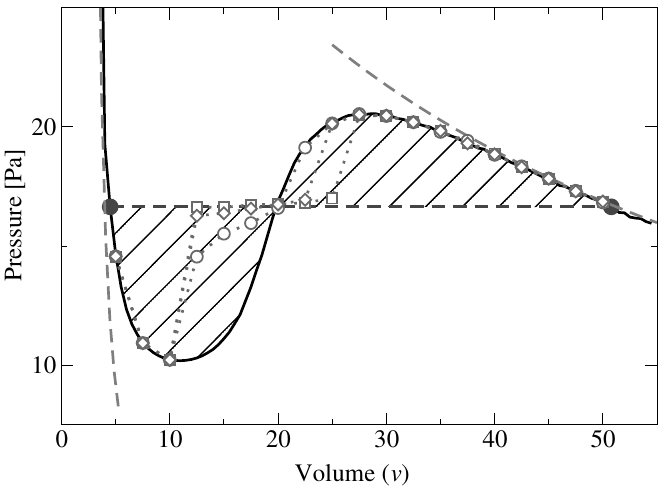}
\caption{Dependence of the pressure, $P$, on the volume per particle, $v$.  The solid line shows the pressure in a small square-base cell with side length $L$=20$d$. In the unstable region, the pressure tends towards the dashed tie line which connects the binodal densities calculated in the large cell ($L$=460$d$). The open circles, squares and diamonds show the pressure calculated in cells of length $L$=80$d$, 120$d$, 160$d$, respectively, demonstrating the convergence to the large-cell limit. The hatched areas illustrate the equal-areas rule. The grey, dashed asymptotes schematically indicate the low temperature branch (left) and high temperature branch (right).}
\label{fig:equal-areas}
\end{center}
\end{figure}

Figure~\ref{fig:equal-areas} shows the dependence of $P$ on the dimensionless volume per particle, $v=V/Nv_p=\bar{\phi}^{-1}$, in a small square-base cell of side $L=20d$ for $A$=2.1$d$ (solid line).
As expected, the pressure exhibits a non-monotonic dependence on the volume, similar to what is observed in a molecular fluid.    However, for the granular system, the pressure curve $P(v)$ is not an isotherm, and the physical origin of its non-monotonic shape is completely different. Unlike a molecular fluid, the granular gas has no attraction between the particles; instead, the dilute phase is heated more effectively due to its intimate, resonant coupling to the vibrating walls, \cite{herminghaus2013,PhysRevLett.107.048002}. As a result, the non-monotonic behaviour in our system can be regarded as a crossover from a low temperature branch at high densities (left dashed curve in Fig.~\ref{fig:equal-areas}) to a high temperature branch at low densities (right dashed curve in Fig.~\ref{fig:equal-areas}). The open symbols indicate the pressure calculated in cells of variable size, showing the convergence to the large-cell limit. We see that the region between the two extrema of $P(v)$ is unstable against phase separation, resulting in a pressure corresponding to two-phase coexistence, $P^{\ast}$ (horizontal dashed line).

\begin{figure}[t]
 \begin{center}
 \includegraphics[clip,width=1.0\columnwidth]{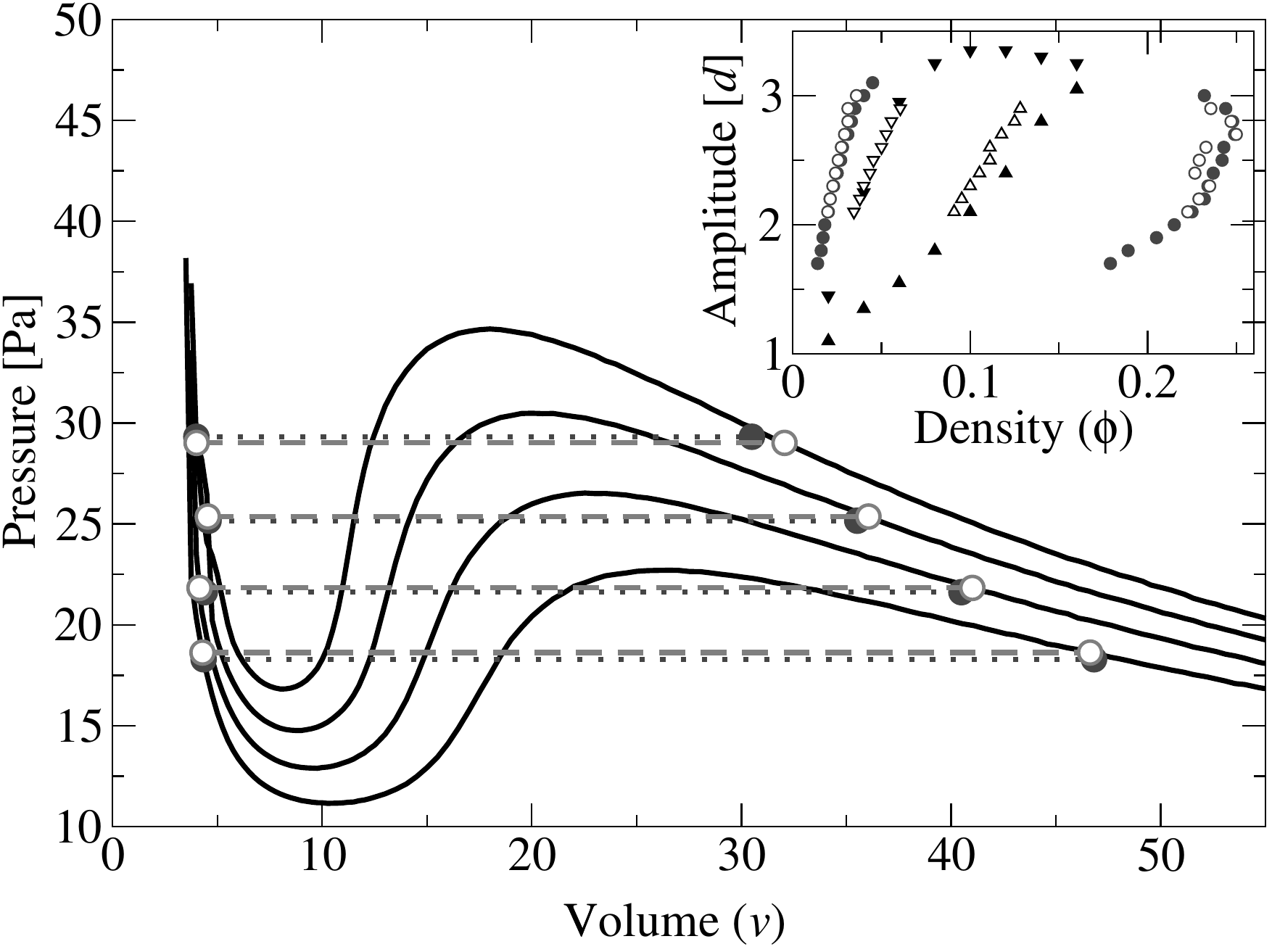}
  \caption{
  Main panel: The solid lines show the pressures calculated in small cells
for driving amplitudes in the range $2.2d{\leq}A{\leq}2.8d$.
The filled circles show the corresponding binodal densities and the pressures calculated in long, thin systems.
The open circles show the pressure and densities predicted using an equal-areas construction.
Inset: Phase diagram for the liquid-gas-like phase separation.  The filled and open circles show the binodal points determined by the long cell, and those predicted by  $P(v)$ and the equal-areas construction, respectively. The triangles show the spinodal points from the long cell and those predicted by the unstable region of $P(v)$, respectively.}
\label{fig:long-cell}
 \end{center}
\end{figure}

\begin{figure}[t]
 \begin{center}
  \subfloat{\includegraphics[clip,width=1.0\columnwidth]{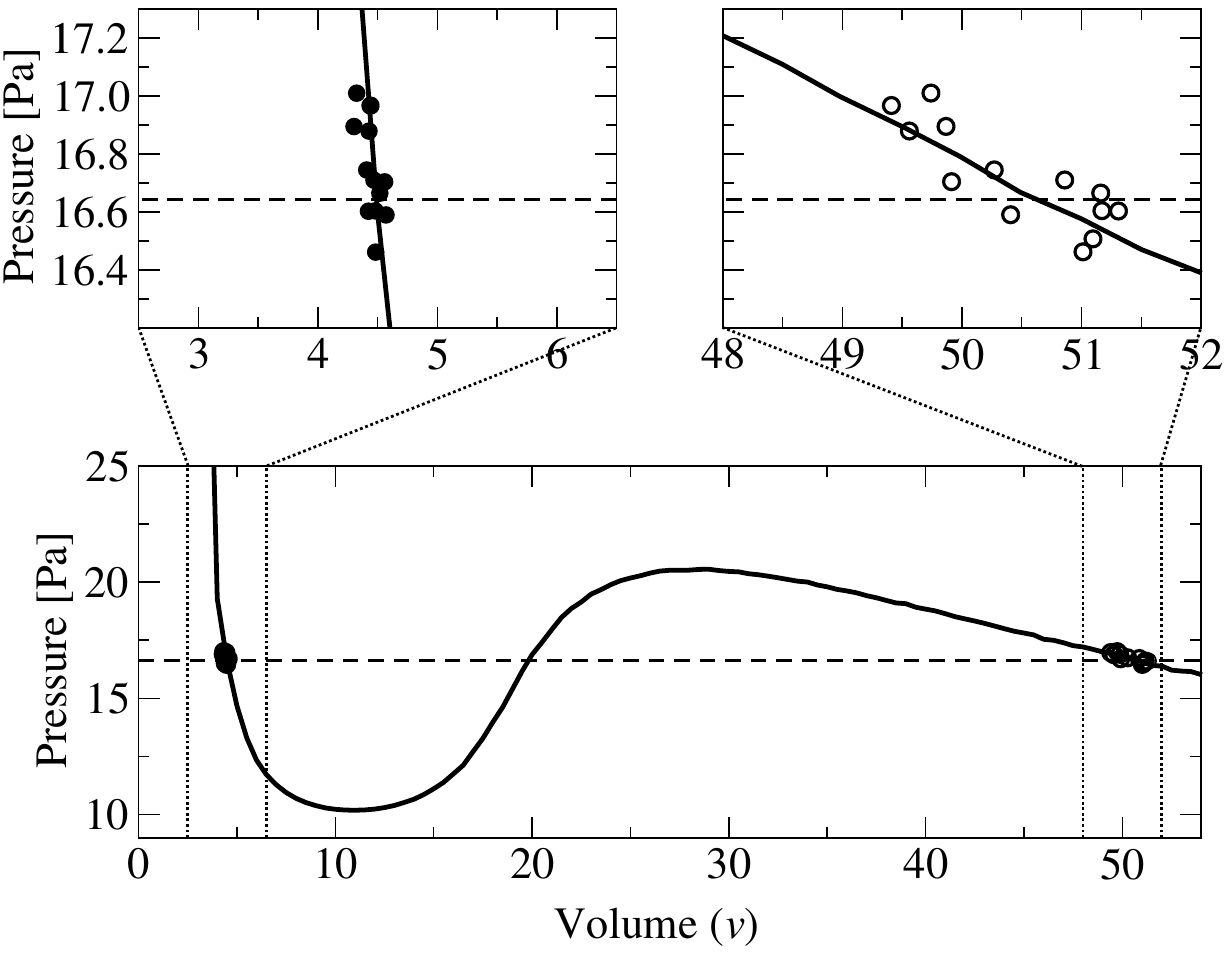}}
  \caption{The pressure-volume curve obtained from the small cell and an expanded view close to $P^*$. The filled and open circles mark the averaged pressure and binodal volumes per particle for the liquid and gas phases respectively, obtained at different times. At each time, the average pressures in the two phases are equal to a good approximation.}
 \label{fig:equation-of-state}
 \end{center}
\end{figure}

Let us now turn to the equal-areas construction. We observe that the horizontal dashed line in Fig.~\ref{fig:equal-areas}, corresponding to $P^*$, creates two approximately equal areas bounded above and below by the curve $P(v)$ (hatched). To investigate this in more detail, the main panel of Fig.~\ref{fig:long-cell} shows $P(v)$ calculated in the small cell for a range of driving amplitudes. For each curve, the closed circles show the pressure determined using the large  cell, the open circles indicate the pressure predicted by the equal-areas construction.  It is quite striking that, for all the amplitudes investigated, the equal-areas construction predicts the pressure and the coexisting densities rather well. 
The inset in Fig.~\ref{fig:long-cell} shows the spinodal and binodal lines determined using the large cell (filled symbols), and the predictions made by using the small cell and equal-areas construction (open symbols). Again, we find remarkably good agreement given that we have no recourse to equilibrium thermodynamics, and as such there is no {\it a-priori} reason to expect the equal-areas construction to hold. 

In order to understand why the coexisting pressure, $P^*$, in the phase-separated system is so close to the equal-areas pressure, $P_{EA}$, obtained from the small-cell simulations, let us first consider a small cell with a fixed number of particles. If such a cell is expanded at some ambient pressure, from a volume corresponding to a mean liquid density $\bar{\phi}=\phi_l$ to that of the gas phase, $\phi_g$, the work done per particle is readily obtained by integration of $P(v)$, since $P(v)$ does not depend on system size. If the expansion is carried out at $P_{EA}$, then the total work done per particle vanishes.  

Consider now a large system with two coexisting phases. Mechanical equilibrium requires the pressure to be equal in both phases. Figure~\ref{fig:equation-of-state} shows the mean pressure, $\bar{P}$, and volume per particle in each phase in the large cell, averaged over a few driving cycles. As the pressure fluctuates around $P^*$, the corresponding densities fluctuate so as to remain on the pressure curve $P(v)$, confirming that $P(v)$ does indeed act as the equation of state for our system.  The pressure fluctuations are found to be tightly correlated with displacements of the interface between the two phases because the number of particles is fixed.   These observations suggest the following simple model which is able to explain the origin of the equal-areas construction.

Let the total volumes of the phases be $V_i=N_iv_i$, where $N_i$ and $v_i$ are the number of particles and the specific volumes, respectively, for $i\in \{l, g\}$.   
Consider a system of a fixed number of particles in a small box at a pressure $\bar{P}$ slightly away from $P^*$, $\bar{P} = P^* + \mathcal{P}$. 
The mechanical work done per particle to expand the box from specific volume $v_l^*$ to $v_g^*$ is, to leading order in $\mathcal{P}$, $w=(v_g^*-v_l^*)\mathcal{P}$. 
Here $v_l^*$ and $v_g^*$ are the corresponding liquid and gas specific volumes at $P^*$. 
For any fluctuation, $\delta N_l+\delta N_g=0$ and $\delta V_l+\delta V_g=0$. The volumes are related to the pressure through $P(v)$  such that, to leading order,
$\delta v_i=-g_i\delta \mathcal{P}$, where $g_i = -\partial v/\partial P\mid_{v=v_i^*}$ is the compressibility. From these relations it follows that, for an additional small change in $\mathcal{P}$,  the change in particle number is $\delta N_g=[(g_lN_l^*+g_gN_g^*)/(v_g^*-v_l^*)]\delta \mathcal{P}$,    
where $N_l^*$ and $N_g^*$ are the number of liquid and gas particles at $P^*$. 
Thus, the work done to move $\delta N_g$ particles across the interface is 
$\delta W_{N}=w\delta N_g=(g_lN_l^*+g_gN_g^*)\mathcal{P} \delta \mathcal{P}$. 
For a finite fluctuation $\mathcal{P}$ the total work is
\begin{equation}
W_{tot}=\int_0^{\mathcal{P}} \delta W_{N}= \frac{1}{2} (g_lN_l^*+g_gN_g^*)\mathcal{P}^2 ,
\label{Eq:Wtot}
\end{equation}
which is quadratic in $\mathcal{P}$. Consequently, any fluctuation that shifts the pressure away from $P^*$ while keeping the volumes per particle on $P(v)$ requires mechanical work to be done.

The energy needed to move the system away from $P^*$ is provided by momentary imbalances between the energy injected by vibration and the energy dissipated in collisions. Because the system is dissipative, any residual mechanical energy can always be released, moving the pressure back towards $P^*$.  This energy imbalance allows the bulk granular temperatures to fluctuate independently of the mechanical work.
Although, in general, the equation of state for a granular system does depend on temperature, in a strongly vibrated system the temperature is fixed by the rate of energy injection and dissipation, Therefore,  the fluctuations of the interface arise only from additional mechanical work done on one phase by the other.

Although in a NESS the average dissipation balances with the injection, energy conservation is strongly violated.  
Thus, it is quite unexpected that a global mechanical-energy minimization argument alone is sufficient to explain the behavior of the system.
This is distinct from that of minimizing the local free energy, as applicable in thermodynamic equilibrium.
In the main panel of Fig. \ref{fig:long-cell} it is clear that the pressure calculated in the long cell deviates slightly 
but systematically from the equal-areas rule (only by less than 3\%).  It can be shown that this inequality follows from the higher order terms neglected in Eq.~(\ref{Eq:Wtot}), see supplemental material.  

Finally it is interesting to quote from Maxwell's discussion of the equal-areas rule in equilibrium systems: {\it ``Since the temperature has been constant throughout, no heat has been transformed into work''}~\cite{Maxwell1875}. In our system the  temperature is not constant throughout an expansion, yet to a good approximation an equal-areas rule still appears to be applicable based on the minimization of the residual mechanical work alone.
It would be interesting to investigate whether this principle can be extended to a wider class of systems driven far from equilibrium. 

\bibliography{equal-areas}

\end{document}